\documentclass[twocolumn]{aa}
\usepackage{natbib}
\bibpunct{(}{)}{;}{a}{}{,} 
\usepackage{graphicx}
\usepackage{txfonts}
\def\hii{\hbox{H~{\sc ii}}}
\def\uchii{\hbox{UCH~{\sc ii}}}
\def\micron{\hbox{$\mu$m}}    
\begin{document}
   \title{The Hot Core--Ultracompact \hii{} Connection in G10.47+0.03
\thanks{Based on observations collected at the European Southern Observatory,
   	La Silla, Chile. Prop.ID:67.C-0359(A) and Prop.ID:69.C-0189(A)}}


   \author{I. Pascucci\inst{1}
          \and
	  D. Apai\inst{1}
	  \and
	  Th. Henning\inst{1}
	  \and
	  B. Stecklum\inst{2}
	  \and
          B. Brandl\inst{3}
	  }

   \offprints{I. Pascucci, \email{pascucci@mpia.de}}

   \institute{
   Max-Planck-Institut f\"{u}r Astronomie,
   K\"{o}nigstuhl 17, D-69117 Heidelberg, Germany 
   \and
   Th\"uringer Landessternwarte Tautenburg, Sternwarte 5, D-07778 Tautenburg, Germany
   \and
   Leiden Observatory, P.O. Box 9513, NL-2300 RA Leiden, The Netherlands
             }

   \date{Received December 16, 2003; accepted June 30, 2004}

   \abstract{
    We present infrared imaging and spectroscopic data of the complex 
    massive star-forming region G10.47+0.03.
    The detection of seven mid-infrared (MIR) sources in our field 
    combined with a sensitive Ks/ISAAC image allows to establish a very 
    accurate astrometry, at the level of 0\farcs3.
    Two MIR sources are found to be coincident with two ultracompact \hii{} regions
    (\uchii s) within our astrometric accuracy.
    Another MIR source lies very close to three other \uchii{} regions and to the 
    hot molecular core (HMC) in G10.47+0.03.
    Spectroscopy of two of the most interesting MIR sources allows to identify the
    location and spectral type of the ionizing sources.
    We discuss in detail the relationship between the HMC, the \uchii{} regions and 
    the nearby MIR source. The nature of the other MIR sources is also investigated.
   }

   \maketitle
%

\section{Introduction}
	Ultracompact \hii{} regions (\uchii) represent one of the youngest 
	detectable stages in the formation of massive stars. 
	They are small, dense and bright radio
	sources still embedded in their natal molecular cloud
	(see \citealt{2002ARA&A..40...27C} for a review). 
	A number of them have been found to be associated with small, hot and dense cores of 
	molecular gas (HMCs, \citealt{2000prpl.conf..299K}). 
	There is evidence that some HMCs are internally heated and centrally 
	dense as expected from cloud core collapse. 
	This has been studied in detail and proved in at least three cases:
	the HMCs in G31.41+0.31 and G29.96-0.02 \citep{1999AAS...195.7307W,2001A&A...371..287M}
	and the well-known HMC in Orion \citep{1998ApJ...497..276K,2002ApJ...574L.163D}.
	The internal heating together with the coincidence of HMCs with maser sites
	would support the idea that HMCs are
	the precursors of \uchii{} regions \citep{1995RMxAC...1..137W}.
	Nevertheless, cases like G34.26+0.15 point towards a completely different scenario
	in which the HMCs would be externally heated by interaction 
	with the nearby \uchii{} regions \citep{1999ApJS..125..143W}.
	
	Very recently a number of groups tried to detect the mid-infrared (MIR) counterparts of
	HMCs aiming to identify the embedded newly born massive star(s) 
	\citep{2002A&A...392.1025S,2002ApJ...564L.101D,2003ApJ...598.1127D,2003A&A...L}.
	These studies demonstrated the importance of establishing a reliable and precise
	astrometry to correctly interpret the MIR data.	 
	Due to the lack of MIR sources with near-infrared/visible counterparts,
	the astrometry of MIR images is often set by the telescope pointing accuracy 
	(typically a couple of arcseconds) or by using the location and morphology of
	radio and/or maser sites whose connection with the MIR emission is yet unknown.
	Up to now, only three out of nine HMC candidates have been detected in the MIR regime
	\citep{2002ApJ...564L.101D,2002A&A...392.1025S,2003ApJ...598.1127D}.	
	
	In this paper we present infrared imaging and spectroscopy of the complex high-mass 
	star-forming region G10.47+0.03 (hereafter G10.47).
	This region represents a promising site to search for the connection
	HMC--\uchii{} regions:
	Four \uchii{} regions have been detected at radio wavelengths 
	(\citealt{1989ApJS...69..831W}, hereafter WC89 and 
	\citealt{1998A&A...331..709C}, hereafter CHWC98), three of them 
	are found to be still embedded in the HMC traced by ammonia and methyl 
	cyanide emission (CHWC98, \citealt{1996A&A...315..565O}). 	
	Many other complex molecular species 
	\citep{1996A&A...307..599O,1998A&AS..133...29H,1999A&A...341..882W}
	and strong maser emission of OH, H$_2$O and CH$_3$OH 
	\citep{1995MNRAS.277..210C,1996A&AS..120..283H,1998MNRAS.301..640W}
	have been detected towards the region of the HMC and the embedded \uchii{} regions.
	Recent interferometric millimeter observations show that the emission of
	a dust clump peaks at the location of the two most 
	embedded \uchii{} regions \citep{2002cdsf.conf...32G}.			
		  	  
	Based on astrometric accuracy at the level of 0\farcs3,
	we aim to study in detail the relationship between the \uchii{} regions,
	the HMC and the MIR emission towards G10.47. 
	In our discussion we adopt a distance of 5.8 kpc for the massive star-forming region,
	as determined by \citet{1990A&AS...83..119C} 
	from their measured ammonia velocity and the rotation curve of \citet{1986PhDT.........9B}.	
	In Sect.~\ref{sect:obs} we describe the observations and 
	data reduction. Two independent methods are considered to establish the astrometric reference 
	frame of our mid-infrared images (Sect.~\ref{astrometry}). 
	The results and their interpretation are presented in Sect.~\ref{sect:results}
	and ~\ref{sect:discussion}.
	The last Section summarizes our findings.
\section{Observations and data reduction} \label{sect:obs}
   	The mid-infrared observations presented in this paper were obtained at different wavelengths
	and with different instruments.
	An additional Ks image was taken at the VLT with the ISAAC camera to calibrate the 
	astrometry of our MIR images (Sect.~\ref{ISAAC}). 
	The 1~$\sigma$ sensitivities of the different observations
	are collected in Table~\ref{tab:sens}.
\subsection{SpectroCam10 imaging}\label{SpectroCam10}
	The first observations were carried out in June 1999 using SpectroCam--10 
	\citep{1993SPIE.1946..334H}, the 10~\micron{} spectrograph 
	and camera for the 5-m Hale telescope\footnote{Observations at the Palomar 
	Observatory made as part of a continuing collaborative agreement 
	between the California Institute of Technology and Cornell University.}.
	SpectroCam--10 is optimized for wavelenghts from  8 to 13\,\micron{} and has a
	Rockwell 128$\times$128 Si:As Back Illuminated 
	Blocked Impurity Band (BIBIB) detector with a plate scale 
	of 0\farcs25 per pixel. 
	In imaging mode the unvignetted field of view is 15\arcsec .
	 
	During the observations, we applied filters with central wavelengths at 8.8, 11.7, and 
	17.9\,\micron{} and bandwidths of 1\,\micron{}. 
	We used the chopping/nodding technique with a chopper throw of 
	20\arcsec{} in north-south direction. The 11.7\,\micron{} field
	has been slightly enlarged by observing at three different positions: 
	the first one centered on the radio source (WC89),
	and the other two positions offset by few arcseconds.
	The total on-source integration time amounts to 5, 8, and 3~minutes for
	the 8.8, 11.7, and 17.9\,\micron{} filters, respectively. 
	The source $\delta$\,Oph has been observed immediately after our target and we use it as
	reference star for the flux calibration  
	(fluxes are taken from \citealt{1999AJ....117.1864C}).
	
	The data reduction was performed using self-developed IDL scripts in
	the standard fashion: the off--source beams of the chopping and nodding
	were used to remove the sky background and its gradients.  	 	
	We also improved our signal-to-noise ratio by applying the wavelet 
	filtering algorithm of \citet{1996A&AS..118..575P}, a 
	flux-conservative method useful to search for faint extended 
	emission.	 
\begin{figure}
	\centering
        \includegraphics[angle=90,width=8cm]{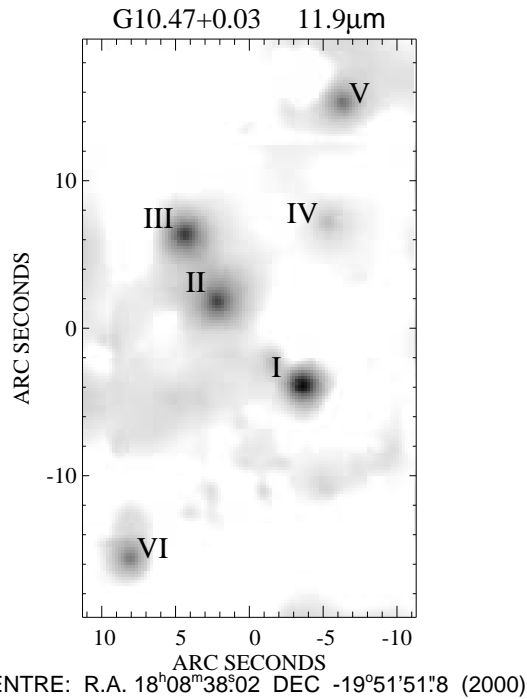}
        \caption[]{TIMMI2 image at 11.9\,\micron{} 
	with source labeling. 
	Only the positive beams are shown in the figure. 
	The wavelet filtering algorithm of \citet{1996A&AS..118..575P}
	has been applied to enhance the image quality. 
	Source VII lies outside the figure.
	At about 35\arcsec{} from the HMC in G10.47, source VII coincides
	with a \uchii{} region in the star-forming region G10.46+0.03
	(see Sect.~\ref{res:ima}).} 
        \label{fig:iden}
\end{figure}
\subsection{TIMMI2 imaging}\label{TIMMI2}	
	Further mid-infrared observations were performed with the
	Thermal Infrared Multimode Instrument TIMMI2 \citep{2000SPIE.4008.1132R}
	mounted on the ESO~3.6m telescope.
	TIMMI2 can operate like a spectrograph and imager in the M(5\,\micron), 
	N(10\,\micron) and Q(20\,\micron) atmospheric bandpasses.
 	The detector is a 320$\times$240 Si:As High-Background Impurity Band 
	Conduction array with a pixel scale of 0\farcs2 in the N and Q 
	imaging modes.	
	
	The observations took place in two periods: In May~2001 we imaged the
	region at 9.8 ($\Delta\lambda$\,= 0.9\,\micron ), 11.9 ($\Delta\lambda$\,= 1.2\,\micron ), 
	12.9 ($\Delta\lambda$\,= 1.2\,\micron ) and 20.0\,\micron{} ($\Delta\lambda$\,= 0.8\,\micron ) with total 
	integration times of 7, 22, 9, and 14 minutes, respectively;
	In March~2003 we obtained deeper images in the 11.9, 12.9 and [NeII]
	($\lambda$\,=\,12.8\,\micron , $\Delta\lambda$\,=\,0.2\,\micron ) 
	filters for a total on-source time of 36, 36, and 30 minutes.
	During the first run, chop (north-south) and nod (east-west) throws of 10\,\arcsec{} have been used.  
	In order to enlarge the field of view and thus
	improve our astrometry (see Sect.~\ref{astrometry}), we applied a larger 
	chop/nod throw of 20\arcsec{} during the observations of March 2003.
	 	
 	The standard stars HD~169916 and HD~81797 have been used to flux calibrate
	the data set from May~2001:
	HD~169916 at the 11.9 and 12.9\,\micron{} wavelengths, while
	HD~81797 for the 9.8\,\micron{} and Q band filters.
	During the second run, we observed the flux calibrator stars HD~123139
	at 11.9\,\micron{} and HD~133774 at 12.9\,\micron{} and in the [NeII] filter 
	immediately before and after our target. 
	The corresponding flux densities of all the standards are taken from 
	\citet{1999AJ....117.1864C}. The data reduction was performed as described in
	Sect.~\ref{SpectroCam10}. The 11.9\,\micron{} image obtained from the reduction
	of the March 2003 dataset is shown in Fig.~\ref{fig:iden}.
\subsection{TIMMI2 spectroscopy}
	N-band spectroscopy of sources II and III (see Fig.~\ref{fig:iden})
	was obtained with TIMMI2 on 31 May 2003. 
	A slit width of 1.2\arcsec{} was applied yielding to a resolving power
	of $\sim$170 at 10\,\micron . 
	We used the standard chopping/nodding technique along the slit with
	a throw of 10\arcsec. 
	The on-source integration time amounted to 43 minutes for object II and 32 
	minutes for object III.
	Both sources were observed at an airmass of $\sim$1.02 during almost 
	photometric conditions. The standard star HD178524 was observed 
	at the same airmass as the targets.

	The data reduction was performed using the IDL 
	pipeline kindly provided by R. Siebenmorgen \citep{2004A&A...414..123S}.
	Smaller modifications were made to speed up the pipeline and
	improve its robustness. 	
	The flux calibration of the spectra was
	tied to the results of our 12.9\,\micron{} measurements with TIMMI2.
   \begin{table*}
      \caption[]{Sensitivities (1~$\sigma$) and angular resolutions.}
         \label{tab:sens}
     $$ 
         \begin{array}{p{0.2\linewidth}p{0.1\linewidth}p{0.1\linewidth}p{0.1\linewidth}l}
            \hline
            \noalign{\smallskip}
            Camera		& $\lambda$ 	& 1~$\sigma$	& Resolution$^a$ \\
                   		& [\micron]   	& [mJy/beam] 	& [\arcsec]\\
            \noalign{\smallskip}
             \hline
            \noalign{\smallskip}
              			& 8.8	    	& 21		&  0.7		\\
            SpectroCam-10	& 11.7		& 9		&  0.6		\\
	    			& 17.9		& 1584		&  0.9		\\
		\noalign{\smallskip}
				\hline  
               	\noalign{\smallskip}
				& 9.8		& 35		&  1.0		\\
	   TIMMI2(May~2001)	& 11.9		& 8		&  0.8		\\
				& 12.9		& 9	        &  0.9		\\
				& 20.0		& 420		&  1.4		\\
		\noalign{\smallskip}
				\hline 
               	\noalign{\smallskip}
	   			& 11.9		& 6		&  0.8		\\
	   TIMMI2(March~2003)	& 12.9		& 4		&  0.9		\\
				& [NeII]	& 11		&  0.9		\\
		\noalign{\smallskip}
				\hline 				 
               	\noalign{\smallskip}
	ISAAC			& 2.16		& 0.017		&  0.6		\\										                
	    \hline
         \end{array}
     $$          
\begin{list}{}{}
\item[$^{\mathrm{a}}$] The angular resolution is given as the FWHM of the observed standard stars
for wavelengths smaller than 11\,\micron{}. For longer wavelengths the images are diffraction limited and the
angular resolution is the first zero of the Airy ring.
\end{list}
   \end{table*}   
\subsection{ISAAC Ks-band imaging}\label{ISAAC}
	We obtained deep Ks-band images using the ISAAC detector mounted
	on the Antu unit (UT1) of ESO's Very Large Telescope at 
	Cerro Paranal, Chile. The observations took place on 20 
	July 2002 in visitor mode in the ESO-programme 69.C-0189(A). 
	The goal of the observations was to establish a reliable astrometric
	reference frame to align the MIR detections. 	
	Therefore only images at the Ks-band were taken. Due to instrumental problems
	the long wavelength arm (ALADDIN Array) was used. To avoid 
	the saturation of the bright stars, the shortest possible integration
	time (0.34s) was applied with a five times repetition at each of 
	the 20 positions. The resulting total observational time was
	approximately 34 seconds. The field of view was $ 2.5' \times 2.5'$.	
	The observations were conducted at an airmass of 1.3 and a (visual) 
	seeing of 0\farcs75.	
	
	We applied the standard near-infrared (NIR) reduction method to the obtained
	data as follows: flat fields, bad pixel masking and 'moving sky' 
	subtraction. 	
	The photometry of the resulting image has been carried out by placing
	an aperture with a radius equal to 2.25\arcsec{} on the individual sources. 
	The flux calibration was performed using the standard star 9149 from
	\citet{1998AJ....116.2475P} which was observed at the beginning of the night.
	We checked the resulting photometry on 10 bright stars also present 
	in the 2MASS database. 
	The estimated error on the photometry is 0.1 magnitude.
\begin{figure*}
        \centering
	\includegraphics[angle=90,width=18cm]{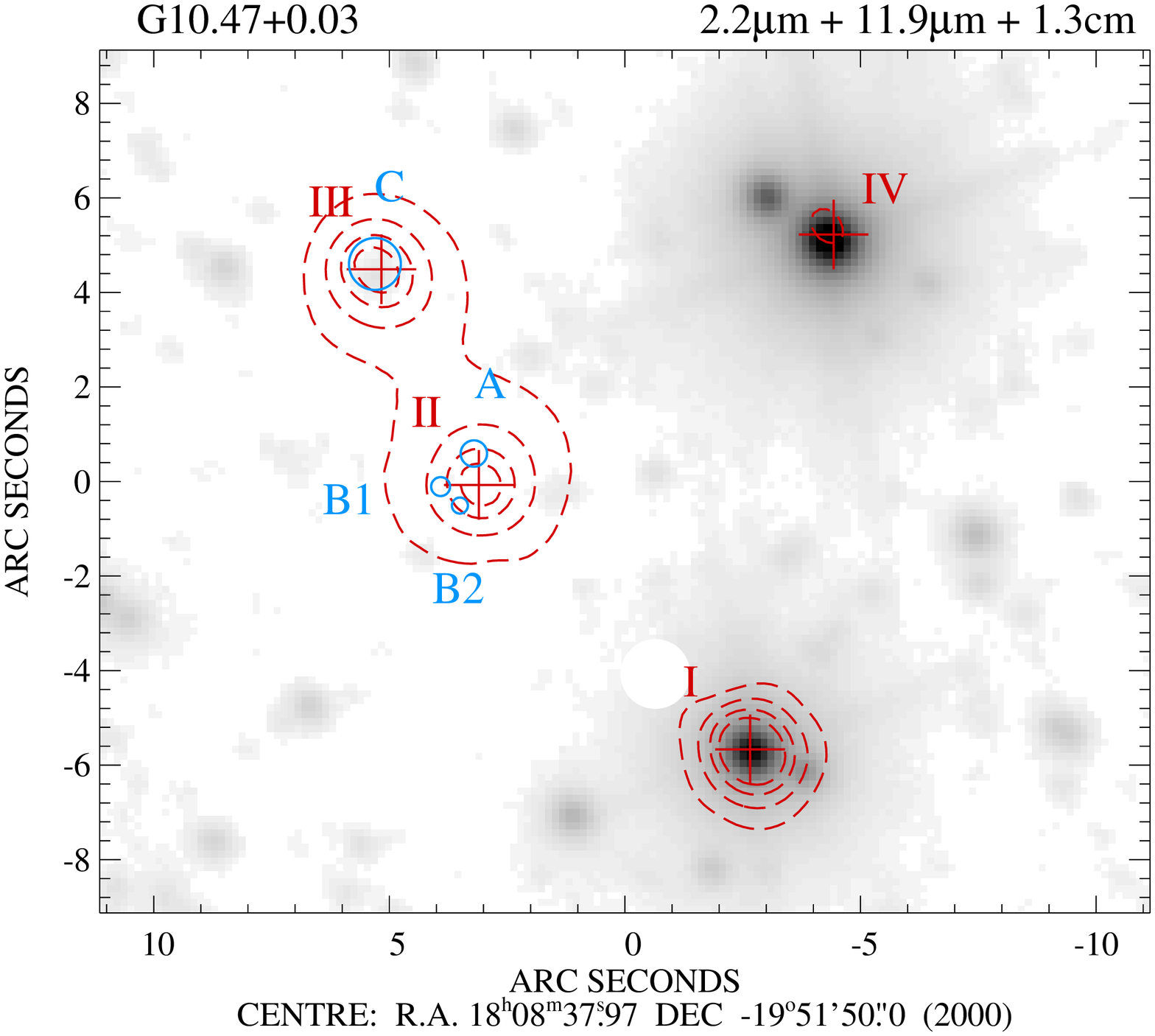}
        \caption[]{TIMMI2 image at 11.9\,\micron{} (dashed red contour 
	in the electronic version)
	superimposed on the ISAAC image (grey scale) at 2.16\,\micron .
	Red contour levels go from 6.25 to 25~mJy/${\sq \, \arcsec}$
	with steps of 6.25 mJy/${\sq \, \arcsec}$
	(source labeling as in Fig.~\ref{fig:iden}).
	The peak positions of the four MIR sources are marked by red crosses.
	The blue empty circles correspond to the four \uchii{} regions,
	namely A, B1, B2, and C, observed by CHWC98 
	in the 1.3\,cm continuum. The diameter of the circles coincides with 
	the full
	width at half power (FWHP$\equiv$FWHM) of the radio emission. 
	The millimeter emission from the HMC peaks at the location of the \uchii{} regions B1 and B2 
	\citep{1996A&A...315..565O,2002cdsf.conf...32G}.
	With a white circular aperture we mask source VII which enters in the field because
	of the adopted chopping/nodding pattern and subsequent image reconstruction procedure.
	As explained in Sect.~\ref{res:ima}, source VII coincides with the \uchii{} 
	region G10.46+0.03A and is located at about 27\arcsec{} (S-W) 
	from source I.
} 
        \label{fig:2.2_11.9_rad}
\end{figure*}
\section{Astrometry}\label{astrometry}	
	An accurate astrometry is essential to understand the relationship 
	between the MIR emission, the HMC and the
	\uchii{} regions. 	
	Interferometric radio continuum maps have good 
	absolute astrometry, usually at the subarcsecond 
	level\footnote{In the case of G10.47 the error is as small as 
	0\farcs1 (Cesaroni private communication)}.
	However, since the connection
	between the radio and the MIR emission is not yet clarified, the MIR astrometry
	should be determined independently from radio measurements.
	
	The detection of up to seven MIR sources in our 11.9\,\micron{} image of
	March~2003 allows us, for the first time, to establish an independent and accurate
	astrometric reference frame for the HMC and \uchii{} regions.
	Two different approaches have been tested:
	One approach (Method~A) is based on the 2MASS Second Release Point Source Catalogue (PSC), 
	the other approach (Method~B) is based on the USNO2 catalogue combined with our Ks-band image. 
	Both approaches provide a consistent astrometry and are briefly described below.	
\subsection{Method A}		
	Three presumably stellar MIR sources, namely I, IV, and V (see Fig.~\ref{fig:iden}), have  
	near-infrared counterparts in the 2MASS 
	PSC\footnote{The 2MASS PSC positions are reconstructed via the Tycho 2 Catalog} 
	and were used to derive the astrometry of our MIR images.
	The standard deviation of the difference between the 2MASS coordinates and the 
	location of the three MIR sources is only 0\farcs12. 
	We compared the 2MASS and USNO2 positions of 24 stars within $2'$ from
	our target to estimate the local positional accuracy of the 2MASS catalogue.
	The standard deviation of the position differences is 0\farcs34.
	Combining the 2MASS-MIR and USNO2-2MASS errors,
	we estimate an accuracy of 0\farcs4 for our 11.9\micron{} image.	
\subsection{Method B}
	In this method, the near-infrared ISAAC image provides the transition from the USNO2 
	catalogue to our 11.9\micron{} image.
	Within the $2.5'$ field of our deep Ks-band image, we identified 16 stars
	which have optical counterparts. 
	Using the USNO2 coordinates we calibrated the astrometry of 
	the ISAAC image with an accuracy of 0\farcs26. 	
	To establish the astrometry of the MIR image we use four objects detected 
	both at near- and mid-infrared wavelengths (sources I, IV, V, and VI in Fig.~\ref{fig:iden}).
	The standard deviation of the differences between the NIR and the MIR coordinates 
	is only 0\farcs13. Thus, the astrometry of our MIR image is accurate within 0\farcs3.
	The central $22''\times17''$ of our Ks-band image together with
	the superimposition of the 11.9\micron{} contour and the location of the \uchii{}
	regions is shown in Fig.~\ref{fig:2.2_11.9_rad}.	
\begin{figure*}	
        \resizebox{\textwidth}{!}{\includegraphics[angle=90,width=8cm]{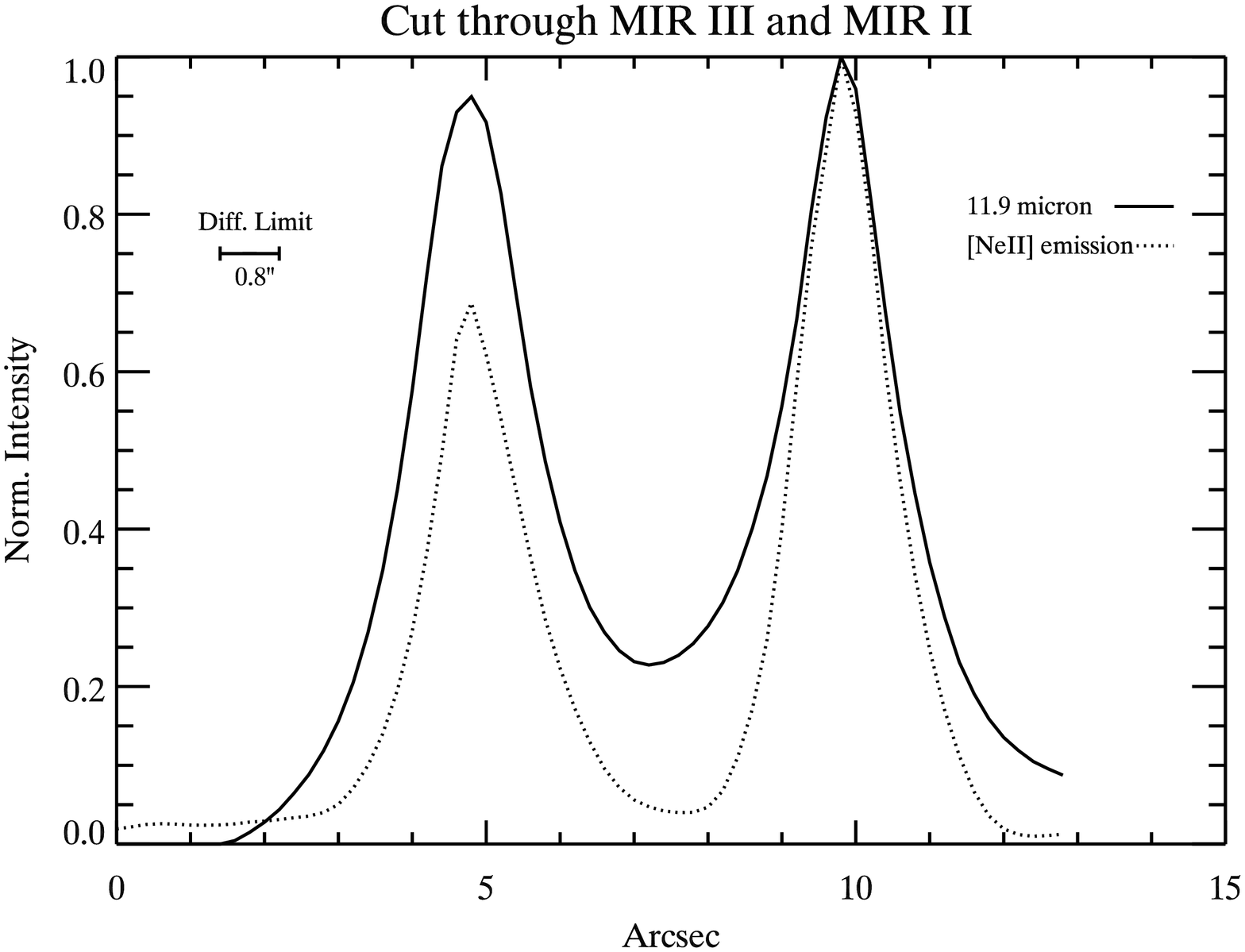}%
        \includegraphics[angle=90,width=8cm]{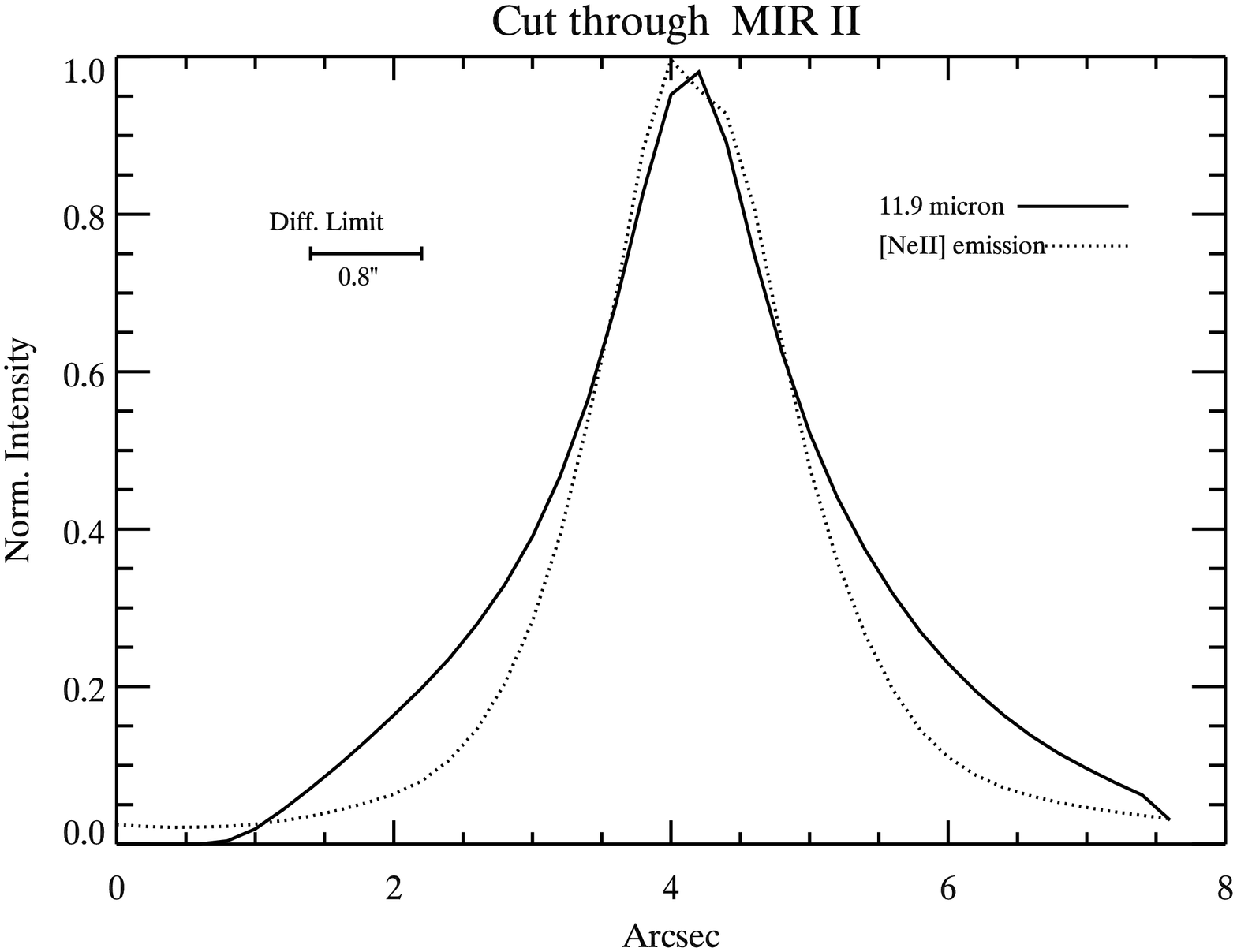}}
        \caption[]{Cuts through the MIR emission of the sources II and III obtained on the
	wavelet-filtered images. The width of the cut is 3 pixels.
	The intensity is normalized to the peak of the MIR source II.
	Left: cut along the line connecting the peaks of the two MIR sources.
	A faint "bridge" of emission is visible between the peaks at 11.9\,\micron .
	Similar faint emission is also present in the non wavelet-filtered images.
	Right: cut in the perpendicular direction for source II showing different sizes at different wavelengths.  
	} 
        \label{fig:cuts}
\end{figure*}
\section{Immediate results} \label{sect:results}
\subsection{Imaging}\label{res:ima}
	Seven MIR sources can be identified in our more sensitive 11.9\micron{} image
	taken during the second TIMMI2 observing run.
	An overview on the designation is given in Fig.~\ref{fig:iden}.
	Five of the MIR sources are also detected at 12.9\micron{} and in the [NeII] filter.
	Only the sources from I to IV are present in the smaller field of the 
	Spectro-Cam observations (see Table~\ref{tab:fluxes}).
	Sources I, IV, V and VI have NIR counterparts and are used
	to determine the astrometric reference frame of our MIR images
	as explained in Sec.~\ref{astrometry}. None of them shows radio emission in the
	1.3~cm (CHWC98) and 6~cm (WC89) continuum maps.
	Source VII is at about 35\arcsec{} south-west of the HMC in G10.47
	and coincides, within our astrometric accuracy, with the \uchii{} region G10.46+0.03A. 
				
	The most interesting MIR sources are II and III because of their vicinity to a group of
	\uchii{} regions (G10.47A, B1, B2, and C) and to the HMC. 
	They appear marginally extended in the [NeII] and in the 12.9\micron{} filters and more extended
	at 11.9\micron . We detect faint emission at this  wavelength between the two sources:
	this "bridge" might be due to the larger extension of the sources
	at 11.9\micron . In Fig.~\ref{fig:cuts} we show a cut along the direction connecting the peak positions
	in two different filters and an additional perpendicular cut for source II.
	
	The \uchii{} region C is found to have a mid- and a near-infrared counterpart in our images
	(see Fig.~\ref{fig:2.2_11.9_rad}): The NIR, MIR and radio peaks  
	are coincident within our astrometric accuracy.
	The 2.16\,\micron{} emission is marginally extended and amounts to 0.22\,mJy.
	Assuming a Gaussian profile both for the source (FWHM~=~0\farcs67) and for the 
	beam (FWHM~=~0\farcs56), we calculate a deconvolved FWHM at 2.16\,\micron{} 
	of 0\farcs4 for source III. The MIR emission is more extended, the  deconvolved source size
	amounts to 1.8\arcsec{} at 11.9\,\micron .
	For comparison, the 1.3~cm radio continuum emission is 0\farcs88 (CHWC98)
	with an extended spherical halo of 3\arcsec{} detected in the lower resolution 6~cm radio map of
	\citet{1993ApJ...418..368G}.
	
	We do not detect NIR emission in the direction of the HMC and the three \uchii{} 
	regions A, B1 and B2. Our NIR upper limit of 0.05\,mJy/beam 
	(3~$\sigma$ sensitivity) translates into a Ks limiting magnitude of 17.8\,mag.  
	Radio measurements predict at least an O9.5 type star as ionizing source of the \uchii{} region A and  
	B0 type stars in the case of the \uchii{} regions B1 and B2.
	The apparent Ks magnitude for an O9.5/B0 star located at the distance of G10.47	
	is 10.3/10.4\,mag (intrinsic infrared colors from \citealt{2000.book.....Allen} and
	absolute visual magnitude from \citealt{1996ApJ...460..914V}).
	Considering the non-detection at NIR wavelengths and the apparent magnitudes calculated above, 
	we find that the extinction towards the HMC and \uchii{} regions is at least 7.4 magnitudes at 2\,\micron . 
	This value translates into a column density of 10$^{23}$\,cm$^{-2}$ when we adopt
	the extinction curve  and the conversion factor to N$_{\rm H}$ from \citet{2001ApJ...548..296W}. 
	For comparison, column density estimates obtained from molecular lines 
	are at least two times larger (\citealt{1996A&A...315..565O}, CHWC98, \citealt{2000ApJ...536..393H}).
	 
	The measured flux densities and peak positions of the seven MIR sources
	are provided in Table~\ref{tab:fluxes}.
	Note that the flux densities related to the [NeII] filter include 
	a contribution from the continuum as well as from the [NeII] line emission.
	The given total fluxes correspond to the fluxes within a synthetic 
	aperture of 4\,\arcsec . 
	The errors of each flux measurement are estimated to be of 15\%
	for the sources from I to IV, and 20\% for sources V, VI and VII. 	
	We report no detection in the 8.8, 9.8, 17.9 and 20\,\micron{} filters.
	In these cases the upper limits for the flux densities can be deduced 
	from the sensitivities given in Table~\ref{tab:sens}.
	 
	We note that the IRAS source 18056-1952, which was considered to be the infrared
	counterpart of the complex massive star-forming region G10.47 (e.g. WC89, 
	\citealt{2000A&A...357..637H}),
	has a 12\,\micron{} flux of 7.93\,Jy, more than 10 times larger than the total flux 
	we measure from the seven MIR sources.
	This, together with the fact that none of our MIR sources is located inside the IRAS
	pointing accuracy ellipse excludes that the IRAS source is related to G10.47.
	For completeness, we mention that the Midcourse Space Experiment (MSX)
	detected a bright unresolved source at 21.3\micron{} whose peak position
	\footnote{The MSX coordinates from \citet{1999STIN...0014854E}  are
	($\alpha_{2000}$;$\delta_{2000}$) 
	= (18$^{\rm h}$08$^{\rm m}$38.38$^{\rm s}$; -19$\degr$51$'$52.6$''$) }
	is at about 3.7\arcsec{} south-east from our MIR source II. 
	The MSX position accuracy is about 2\arcsec{} both in right ascension and declination
	and the flux measured in
	the MSX beam amounts to  22$\pm$1\,Jy. 	
	The large 18\arcsec{} MSX beam includes all our MIR sources but source VII.
	However, considering the steep rise in the spectral energy distribution (SED) of HMCs 
	\citep{1999ApJ...525..808O}
	we expect that the HMC in G10.47 contributes most of the flux at 21.3\micron .
  \begin{table*}
      \caption[]{Measured flux densities and peak positions of the 
      		MIR sources (source labeling as in Fig.~\ref{fig:iden}). 
		Source VII belongs to the star-forming region G10.46+0.03, see Sect.~\ref{res:ima}
		for more details.}

         \label{tab:fluxes}
     $$ 
         \begin{array}{p{0.08\linewidth}p{0.11\linewidth}p{0.11\linewidth}p{0.1\linewidth}p{0.1\linewidth}p{0.1\linewidth}p{0.1\linewidth}l}
            \hline
	    \noalign{\smallskip}
       	 Source  & \multicolumn{2}{c}{\rm{Peak~Position}} &
	 F$_{{\rm 11.7}}$$^{\mathrm{a}}$&
	 F$_{{\rm 11.9}}$$^{\mathrm{b}}$&
	 F$_{{\rm 12.9}}$$^{\mathrm{b}}$&
	 F$_{{\rm [NeII]} }$$^{\mathrm{b}}$	\\
	    \noalign{\smallskip}
            ID       & $\alpha(2000)$ & $\delta(2000)$ & & & &   \\
	           &    [h m s]     &[$^\circ$ $'$ $''$] & 		    
           [mJy]   & [mJy]    & [mJy]   & [mJy] \\
            \noalign{\smallskip}
             \hline
            \noalign{\smallskip}
              I   &18 08 37.78  & -19 51 55.7 & 101& 170& 154 & 220  \\
              II  &18 08 38.19  & -19 51 50.0 & 140& 136& 457 & 943  \\
              III &18 08 38.34  & -19 51 45.6 & 131& 137& 236 & 703  \\     
              IV  &18 08 37.66  & -19 51 44.8 & 23 & 30 & $<$12& $<$33   \\  
	      V   &18 08 37.59  & -19 51 36.6 & --- & 107& $<$12& $<$33   \\
	      VI  &18 08 38.61  & -19 52 07.3 & --- & 94 & 201 & 424   \\ 
	      VII &18 08 36.47  & -19 52 14.9 & --- & 89 & 160 & 309   \\
	    \hline
         \end{array}
     $$ 
\begin{list}{}{}
\item[Note.] The flux densities are computed in the broad-band 11.7, 11.9 and 12.9\,\micron{} filters and in the
narrow-band [NeII] filter.
The estimated flux uncertainty is 15\% for sources from I to IV and 20\% for sources V, VI and VII
\item[$^{\mathrm{a}}$] Symbol "---" indicates that sources V, VI and VII are outside the SpectroCam field
\item[$^{\mathrm{b}}$] Flux densities from the second TIMMI2 run, upper limits represent 3~$\sigma$ sensitivities.
\end{list}
\end{table*}
\subsection{MIR spectroscopy}\label{MIRspec}
\begin{figure*}	
        \resizebox{\textwidth}{!}{\includegraphics[angle=90,width=8cm]{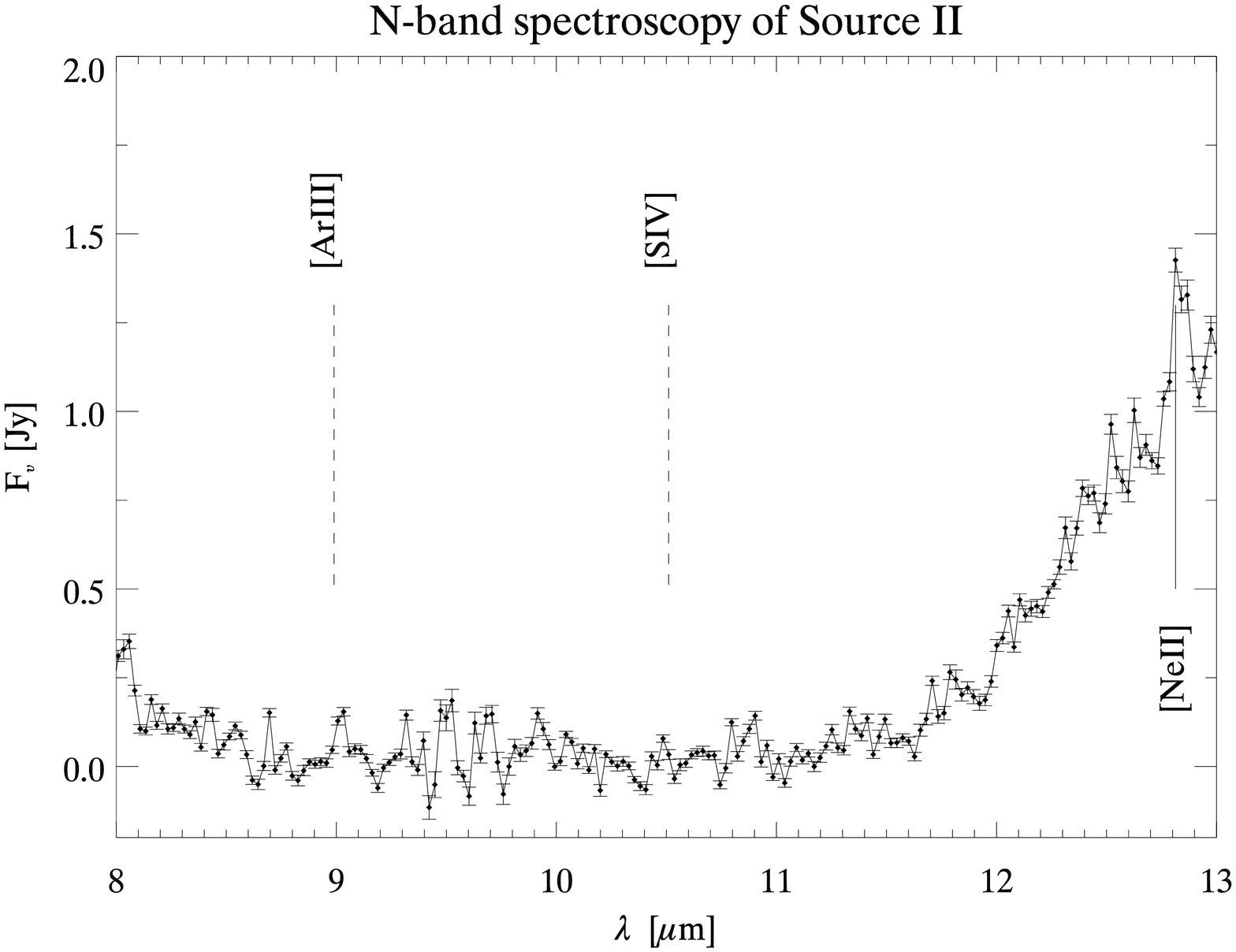}%
        \includegraphics[angle=90,width=8cm]{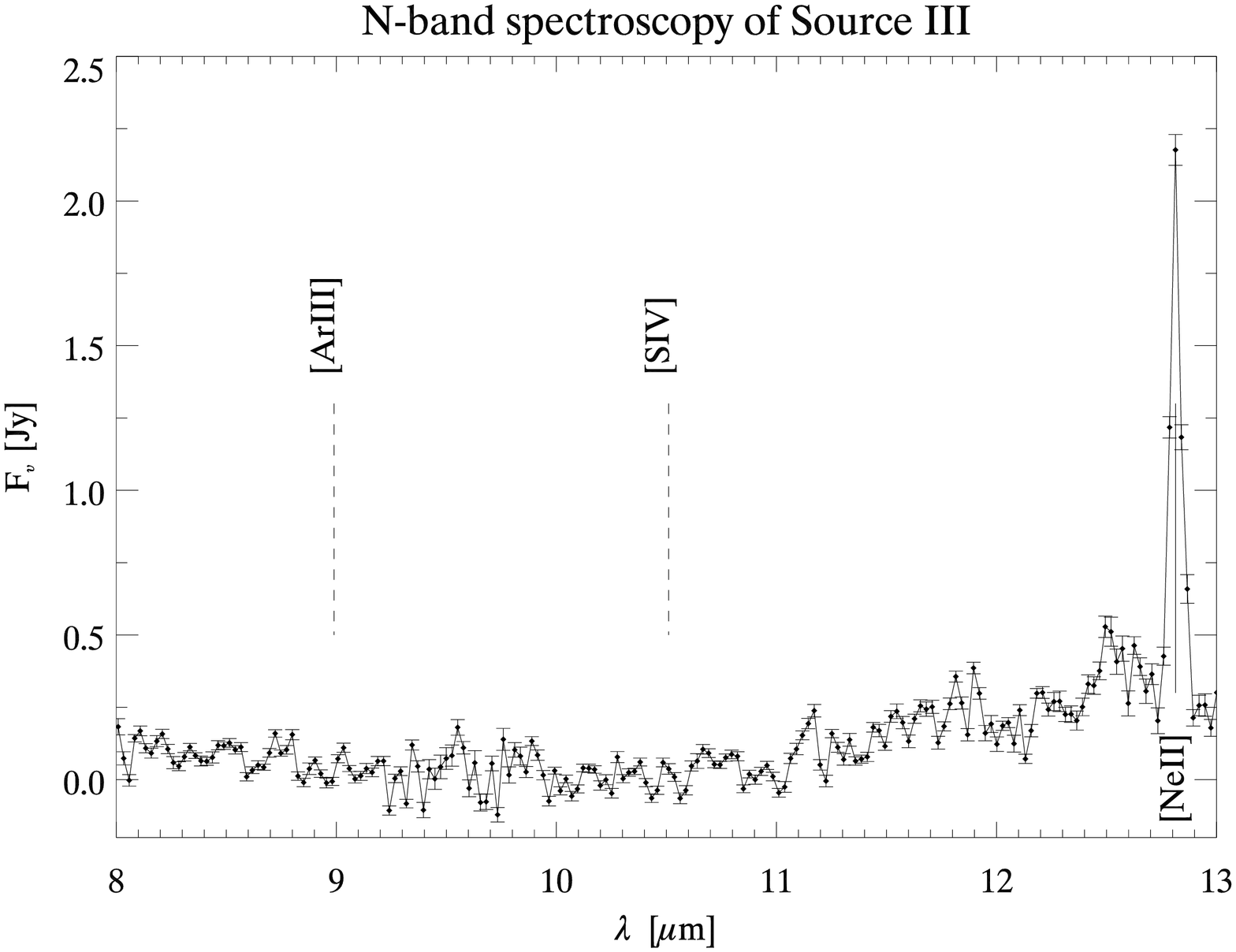}}
        \caption[]{N-band spectroscopy of source II and III.
	The fine-structure emission lines from metal ions
	are marked with solid lines (detection) and dashed lines
	(non detection).} 
        \label{fig:sp}
\end{figure*}

	The calibrated N-band spectra of sources II and III are shown in Fig.~\ref{fig:sp}. 
	A reliable removal of the atmospheric lines is reached in the
	wavelength range 8--13\,\micron .	
	The error plotted on each point represents the flux scatter
	due to photon noise, additional noise may come from the improper subtraction of telluric lines. 	
	
	The spectrum of object II is clearly dominated by the
	silicate absorption feature.  
	The [NeII] emission line at 12.81\,\micron{}
	is detected with a line-to-continuum ratio of about 1.5.
	In contrast, the spectrum of source III shows  strong [NeII] line emission
	and weak silicate absorption.
	None of the other MIR fine-structure lines, like [ArIII] at 8.99\,\micron{} and
	[SIV] at 10.51\,\micron , can be identified in the spectra.
	Following \citet{2001ApJ...553..254O} and \citet{2003ApJ...584..368O}, we estimate
	upper limits for the ionizing stars from the flux ratio of the non-detected
	[ArIII] and [SIV] and the [NeII] emission line.
	Upper limit fluxes for the [ArIII] and [SIV] lines
	are obtained by integrating the de-extincted spectra (see Sec.~\ref{sourceIII}
	for the procedure) in a 0.06\,\micron{}  interval centered on 8.99\,\micron{} and
	10.51\,\micron , respectively. 
	For source II we estimate an O7 type star, while for source III we obtain an O9 
	type star as upper limits. 
	We note that this method is model-dependent: The observed flux ratios
	are compared to CoStar calculations by \citet{1997A&A...322..615S}.
	In the next section, we apply a different method which only depends
	on the observed [NeII] flux to estimate the Lyman continuum flux and thus to determine the 	
	spectral types of the ionizing sources. 		
\section{Discussion on the individual sources} \label{sect:discussion}
	In the following, we discuss the MIR sources with special
	attention on sources II and III, which are close to the \uchii{} regions
	and the HMC in G10.47.
\subsection{Source III and the ultra-compact \hii{} region C}\label{sourceIII}
	Source III is the mid-infrared counterpart of the \uchii{} region C (WC89). 
	Further radio measurements (\citealt{1993ApJ...418..368G} and
	CHWC98) 
	show that a B0 or earlier type star is responsible for its ionization.
	We also detect its NIR counterpart in our Ks-band image
	(see Sect.~\ref{res:ima}). 
	This slightly elongated counterpart has a total flux of 0.22\,mJy at 
	2.16\,\micron , which corresponds to a Ks magnitude of 16.2.	
	The apparent K magnitude for a B0 star at the distance of G10.47
	is 10.4\,mag (intrinsic infrared colours from \citealt{2000.book.....Allen} and
	absolute visual magnitude from \citealt{1996ApJ...460..914V}).
	Thus, we calculate an extinction $A_{2.2}$ of 5.8\,mag at 2.2\,\micron .
 	However, this value only represents a lower limit: part of the 
	observed emission at this wavelength may come from scattered light in a non-spherical
	configuration and from warm dust surrounding the \uchii{} region.
	The presence of warm  dust around the \uchii{} region is evident at longer
	wavelengths where the source appears more extended 
	($\sim$\,1.8\arcsec{} at 11.9\,\micron{} after deconvolution with the
	diffraction-limited beam) and has a flux more than 500 times larger than
	the extincted black body emission from the star.
	
	To obtain a more accurate value for the extinction and to estimate the spectral type
	of the ionizing source independently from the radio measurements, 
	we use our MIR spectrum. 
	We assume a simple model in which the MIR  flux
	from warm dust is described by a black body of temperature T$_{\rm{d}}$
	as optically thick emission and is extinguished by a screen of 
	cold dust in the foreground.	
	For sake of simplicity we do not distinguish between the cold dust in the 
	interstellar medium and that in the molecular cloud. 
	With these assumptions, the observed flux density can be expressed  as:
\begin{equation}\label{eq:flux}
F_{\nu} =  \Theta^2 \, B_{\nu}(T_{\rm{d}}) \, exp(-\tau_{\nu})
\end{equation}	
	with  $\Theta$ being the angular diameter of the source 
	of warm dust, $T_{\rm{d}}$ the temperature of the 
	emitting dust  and $\tau_{\nu}$ the optical depth of the cold dust layer.
	The optical depth  $\tau_{\nu}$ is proportional to the line-of-sight 
	column density $N_{\rm{(H+H_{2})}}$ through the dust extinction 
	cross section per hydrogen nucleon C$_{\rm ext}(\nu)$. 
	These latter values are taken from \citet{2003ARA&A..41..241D}, who recently revised
	the grain size distributions and dielectric functions used to produce the previous
	synthetic extinction curves \citep{2001ApJ...548..296W}.
	We consider two extinction laws for the local Milky Way, one typical for diffuse regions
	($R_{\rm{V}}$=3.1) and one for sightlines intersecting clouds with larger extinction 
	($R_{\rm{V}}$=5.5).
	The continuum of the observed spectrum (7.9--12.75\,\micron{} and 12.9--13.1\,\micron )
	is fitted by using equation~\ref{eq:flux} with $b$, $T_{\rm{d}}$, 
	and $N_{\rm{(H+H_{2})}}$ as free parameters.
	A robust least-square minimization using the Levenberg-Marquardt method 
	is performed to derive the best fit parameters and their
	1$\sigma$ uncertainties which are computed from the covariance matrix.
	The two extinction laws provide the same results within the errors, the best-fit 
	parameters are summarized in Table~\ref{tab:best}.
	From the estimated line-of-sight column density we derive an
	extinction of 4 magnitude at 9.8\,\micron . This value 
	translates into $A_{2.2}$\,=\,6.7 magnitude, about 1 magnitude greater than
	the extinction estimated from the apparent Ks magnitude and the expected flux of a B0 star.
	This means that about half of the measured Ks emission is photospheric flux from the
	ionizing star, the remaining originates from dust re-emission.
	
	To estimate the spectral type of the ionizing source we will use the
	de-extincted flux in the [NeII] line ($I_{12.8}^0$) and a formula which
	links this quantity to the number of ionizing photons per second ($Q_0$).
	The determination of the de-extincted flux in the [NeII] line is done as follows:
	First, we subtract the fitted continuum from the spectrum, then we de-extinct
	the [NeII] flux density by the measured $\tau_{\nu}$, finally we compute 
	the integral in the [NeII] line between 12.75 and 12.9\,\micron{} and 
	the error on the integrated flux as the sum of the errors at each sampled 
	wavelength.	
	The formula connecting $I_{12.8}^0$ and $Q_0$ is derived assuming
	a spherically symmetric, optically thin (CHWC98),
	homogeneous, ionization-bounded \hii{} region.
	The intrinsic flux of an emission line as a function of the radio continuum
	is given by equation~(8) of \citet{2003A&A...407..957M}.
	Equation~(5) from \citet{2003A&A...407..957M}
	provides the number of Lyman continuum photons as a function of the radio continuum.
	The ratio of these two equations allows to express $Q_0$ as a function of the
	de-extincted flux in the [NeII] line:
\begin{equation}\label{eq:ion}	
Q_0 = 2.566\times10^{34} \, T_{\rm e}^{-0.8} \, D^2 \, \frac{I_{12.8}^0}{(Ne^+/H^+)\,\epsilon_{12.8}} \, \, [s^{-1}]
\end{equation}
        In equation~2, $T_{\rm e}$ is the electron temperature in K, $D$ the source distance in kpc,
	$I_{12.8}^0$ the de-extincted flux density of the [NeII] line in erg\,s$^{-1}$\,cm$^{-2}$ 
	and $\epsilon_{12.8}$ the line emissivity in erg\,s$^{-1}$\,cm$^3$. 
	The term $(Ne^+/H^+)$ is the abundance of ionized neon over ionized hydrogen.
	The absence of [ArIII] and [SIV] lines in our spectrum shows that the nebula
	is of low-excitation and therefore we can further assume $(Ne^+/H^+) \sim (Ne/H)$.
	The abundance of Ne relative to H is calculated for the specific galactic distance
	of G10.47 from the equation given in \citet{2003A&A...407..957M}. For a galactocentric distance
	of 2.5\,kpc we compute a value of $2.2\times10^{-4}$ for $(Ne/H)$, which is about two times
	the solar abundance \citep{1998SSRv...85..161G}.
	The calculation of the line emissivity $\epsilon_{12.8}$ is done following the
	formulation of \citet{1974agn..book.....O} (equations 3.25, 4.12 and 5.29) and assuming atomic data from
	\citet{1994A&AS..107...29S}.
	We obtain a value of  4.6$\times$10$^{-23}$ erg\,s$^{-1}$\,cm$^3$ for a 
	$T_{\rm e}$ of 10000\,K and an electron density of $4.2\times10^4$\,cm$^{-3}$ as given by CHWC98.
	The resulting number of Lyman continuum photons with the uncertainty estimated from the error in the [NeII] flux
	and from the two extinction curves is provided in Table~\ref{tab:best}.	
	Our value is about three times larger than that estimated by CHWC98 from the measured radio continuum emission.
	
	Determining the spectral type of the ionizing star from the $Q_0$ parameter requires a careful consideration of
	the recent results from O-star atmosphere models.
	\citet{2002A&A...382..999M} demonstrated that a proper treatment of non-LTE line blanketed atmospheres
	shifts the effective temperature scale of O stars towards lower values for a given spectral
	type in comparison to simpler LTE approaches \citep{1996ApJ...460..914V}.
	Recently, \citet{2004A&A...419..319M} compared the output of different O-star models and concluded that
	the relation between $Q_0$ and $T_{\rm eff}$ is largely model independent. 
	Based on these two results, we adopt the following procedure to derive the spectral
	type of the ionizing source: 
	First we choose the $T_{\rm eff}$ that matches our computed Log($Q_0$) from 
	Table~3 of \citet{1997A&A...322..598S} and then we
	link the temperature to the spectral type using Fig.~1 from \citet{2002A&A...382..999M}.
	In the case of source III the estimated number of Lyman continuum photons is slightly lower
	than the lowest value given in Table~3 of \citet{1997A&A...322..598S} which implies 
	an effective temperature below 32000\,K. 
	These temperatures translate into a B0 spectral type which is in good agreement
	with the spectral type estimated by CHWC98 based on the old temperature scale
	of \citet{1973AJ.....78..929P}.
        In order to check the reliability of our approach, we also compute the radio flux at 1.3\,cm 
	that we expect from equation~(8) of \citet{2003A&A...407..957M}
	and we compare it with the 1.3\,cm flux obtained from CHWC98. Assuming the same values as above and
	the de-extincted flux in the [NeII] line provided in Table~\ref{tab:best}, 
	we calculate a flux of 154$\pm$10\,mJy at 1.3\,cm. This value is about five 
	times larger than that given in Table~3 by CHWC98.
	One reason for this discrepancy might be that some flux is lost in the high resolution VLA configuration. 
	To prove this we compared two measurements at 6\,cm by WC98 and \citet{1993ApJ...418..368G} 
	obtained with two different angular resolutions (about 0\farcs5 and 4.7\arcsec{} respectively). 
	This comparison shows that
	the flux from \citet{1993ApJ...418..368G} is about three times larger than that from WC98 at 6\,cm, thus 
	proving that the more extended VLA configuration indeed overlooked some of the flux from large-scale structures.
        If a similar flux ratio would have been lost in the 0\farcs4 resolution map of CHWC98 at 1.3\,cm,
	the radio flux estimated from our [NeII] line would be only 1.7 time the expected radio flux at 1.3\,cm.
	Given the simple approach we used here, this match is a reasonable good agreement.
	
	Our infrared images and spectrum show that the B0 star ionizing the \uchii{} region C 
	is surrounded by dust which is optically thick at these wavelengths.
	Thus, we cannot use the infrared fluxes to estimate the amount of dust
	surrounding the \uchii{} region. 
	However, we can derive upper limits for the dust mass from the BIMA non-detection at 1.4~mm
	\citep{2002cdsf.conf...32G}. In the natural weighted BIMA map the 1\,$\sigma$ detection
	limit is 20mJy/beam 
	for a beam of 1.98\,\arcsec $\times$1.27\,\arcsec{} (Wyrowski priv. comm.).
	We assume optically thin emission at 1.4~mm, two values for the dust temperature
	(30 and 50~K) and the mass absorption coefficient of the dust from \citet{1994A&A...291..943O}
	for a gas density of 10$^5$\,cm$^{-3}$.
	A dust mass between 0.2--0.4~M$_{\sun}$ is derived. 
	This value translates into a total mass of 20--40\,M$_{\sun}$ if a gas/dust mass ratio of 100 is adopted.
	The corresponding upper limit column densities are 1.9--3.5$\times$10$^{23}$~cm$^{-2}$ for the molecular hydrogen
	(see eq.~2 and 3 of \citealt{2000A&A...353..211H}), well in agreement with the value 
	we estimate by fitting the MIR continuum 
 	of our N-band spectrum (Table~\ref{tab:best}).
	
  \begin{table*}
      \caption[]{Best fit parameters to the N-band spectra and estimated spectral types for the ionizing sources}
         \label{tab:best}
     $$ 
         \begin{array}{p{0.08\linewidth}p{0.16\linewidth}p{0.1\linewidth}p{0.1\linewidth}l}
            \hline
	    \noalign{\smallskip}
       	 	Parameter$^a$  	&	Unit  		      & Source II    &Source III \\
            \noalign{\smallskip}
             \hline
            \noalign{\smallskip}
	    $T_{\rm d}$ 	& 	(K)                   & 271$\pm$19   & 324$\pm$47 \\
	    $N_{\rm (H+H_2)}$\, &($\times 10^{22}$\,cm$^{-2}$) &  23$\pm$2 & 8$\pm$1 \\
	    $\Theta$	& (mas)			      &  29$\pm$16            & 5$\pm$4 \\	
	    $\tau_{9.8}$ 	&                    	      &  9          & 4 \\
	    $I_{12.8}$ 	        &($\times 10^{-13}$\,erg\,s$^{-1}$\,cm$^{-2}$) & 4.9$\pm$0.8     & 19$\pm$1  \\
	    $I_{12.8}^0$        &($\times 10^{-12}$\,erg\,s$^{-1}$\,cm$^{-2}$) & 9$\pm$2          & 7.0$\pm$0.4 \\
	    log(Q$_{\rm 0}$)    &(s$^{-1}$)		      & 47.7$\pm$0.1	     &  47.5$\pm$0.1 \\
	    $T_{\rm eff}$       &   (K)                       & 32060             &  $<$32060  \\
	    SpTy	        &			      & O9.7 & B0 \\
	    \hline
         \end{array}
     $$ 
\begin{list}{}{}
\item[$^{\mathrm{a}}$] 
$T_{\rm eff}$ is determined from Table~3 of \citet{1997A&A...322..598S}, while the spectral type (SpTy)
 from Fig.~1 of \citet{2002A&A...382..999M}.
\end{list}
   \end{table*}  
\subsection{Source II and its relation with the hot molecular core}
\begin{figure}
        \centering
	\includegraphics[angle=90,width=10cm]{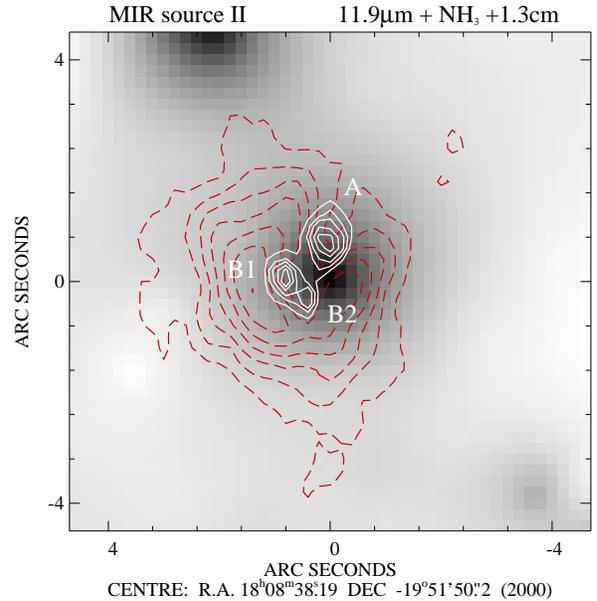}
        \caption[]{Image of the MIR source II at 11.9\,\micron{}
	with contours of the 1.3\,cm radio continuum (solid white lines) and of 
	NH$_3$(4,4) line emission (red dashed lines in the electronic version) from 
	CHWC98.
	White contour levels are spaced by 7\,mJy/beam
	starting from 5 to 33\,mJy/beam, red-dashed contours are spaced by 
	4\,mJy/beam from 3 to 27\,mJy/beam. 
	} 
        \label{fig:11.7_radc_amm_det}
\end{figure}

        Source II is the most exciting among our MIR detections.
	Three \uchii{} regions and a HMC have been identified 
	by CHWC98 in this region.         	
	CHWC98 suggest a picture in which the three massive stars,
	at the stage of \uchii{} regions, are embedded in the dense molecular core 
	traced by ammonia emission.
	Since the \uchii{} regions B1 and B2 are more compact than A and
	show more absorption in the NH$_3$(4,4) transition,  CHWC98
	propose that A is lying closer to the surface of the HMC traced in ammonia, i.e in 
	a less dense region.
	The fact that the three \uchii{} regions are tracing embedded stars is supported by
	the compactness of their radio emission: The two VLA 6\,cm maps from WC89 and 
	\citet{1993ApJ...418..368G}
	at different resolution provide the
	same total flux for the three \uchii{} regions. 
	To show the complexity of this region we superimpose the radio, ammonia and
	mid-infrared emission in Fig.~\ref{fig:11.7_radc_amm_det}.
	
	Up to now, only few cases are known in which MIR emission is detected from the close
	vicinity of HMCs \citep{2002ApJ...564L.101D,2003ApJ...598.1127D}. 
	Proving that such emission arises from the HMC itself would
	support the idea of internally heated cores.
	In Sect.~\ref{astrometry} we demonstrated that our astrometry is accurate to 0\farcs3
	based on stars with counterparts at different wavelengths.
	This high accuracy allows to explore the relation between the \uchii{} phenomena,
	the HMC and the MIR emission.
	Here, the main question is whether the MIR emission 
	 is indicating the presence of a protostar, or of a young massive star 
	 prior to the \uchii{} phase, or arises from dust surrounding the \uchii{} regions.
	To answer this question we use both the mid-infrared images and the spectroscopic data.
	
	Before investigating the  three scenarios, we briefly summarize the
	main observational results: \\
	I) Based on our astrometry, none of the \uchii{} regions coincides with the mid-infrared
	peak of source II:
	this peak is located south of the \uchii{} region A,
	at about the same distance ($\sim$0\farcs6\,=\,3480\,AU) from A and B2. 
	The \uchii{} region B1 is further away from source II at about 0\farcs8.
	The MIR emission does not coincide with the center of the HMC, which is believed
	to be close to the most embedded \uchii{} regions B1 and B2. \\
	II) The MIR emission is extended in the NW-SE direction even in the [NeII] filter
	(deconvolved FWHM of 0\farcs9 after the continuum subtraction). 
	The MIR extension is even larger at 11.9\micron , about 1.8\arcsec{} 
	(see Fig.~\ref{fig:cuts} and Sect.~\ref{res:ima}). \\ 
	III) The MIR spectrum does not show the forbidden lines 
	of [ArIII] and [SIV] thus implying an O7 upper limit for the spectral type of 
	the ionizing source (see Sect.~\ref{MIRspec}). \\
	IV) The combination of PdBI data with IRAM 30-m maps in the CH$_3$CN(6-5) line led
	\citet{1996A&A...315..565O} to identify an extended halo around a 
	compact core. The compact core well matches the location and extension of
	the ammonia HMC (CHWC98). Assuming an abundance of CH$_3$CN relative to H$_2$ of
	$\sim$10$^{-8}$ \citep{1998A&AS..133...29H,1993A&A...276..489O}
	 the column densities of Table~3 from 
	\citet{1996A&A...315..565O} translate into H$_2$ core and halo column densities of
	$\sim$6$\times$10$^{24}$\,cm$^{-2}$ and $\le$3.6$\times$10$^{23}$\,cm$^{-2}$.
	The upper value\footnote{While a CH$_3$CN/H$_2$ ratio of 10$^{-8}$ is representative for the hot core,
	a somewhat lower value is expected for the halo. Currently no such quantitative measurements are 
	available.} 
	for the halo column density agrees well with
	the column density we measure towards source II
	from fitting our MIR spectrum (see Table~\ref{tab:best}).
	However, considering the different techniques and angular resolutions of the observations, 
	we cannot determine the exact line-of-sight location of source II in respect to the HMC. \\
	V) The slope of the radio emission as given from the measurements at 6\,cm (WC89) and 1.3\,cm 
	(CHWC98) is $S_\nu \propto \nu^{0.9}$ for the \uchii{} region A and $S_\nu \propto \nu^{1.4}$ for B.
	Both slopes are  different from that expected in the case of
	ionization due to a spherical stellar wind ($S_\nu \propto \nu^{0.6}$, \citealt{1975A&A....39....1P}).
	Also the size of the \uchii{} region A increases towards shorter wavelengths,
	contradicting the wind hypothesis 
	($R_{\nu}\propto \nu^{-0.7}$, \citealt{1975A&A....39....1P}).

	We now consider the first two scenarios, namely a massive protostar 
	or a massive star embedded in the HMC at a stage prior to the \uchii{} phase.
	In the  protostellar phase, accretion provides most of the luminosity
	and the Lyman continuum emission is expected to be much fainter than for a ZAMS O-B star.
	This stands in evident contradiction with the observed presence of the [NeII] emission from 
	source II, effectively excluding its massive protostar nature.
	Furthermore, the detection of  [NeII] emission from source II implies a radio continuum emission as 
	large as 265$\pm$59\,mJy at 1.3\,cm ($I_{12.8}^0$ from Table~\ref{tab:best}
	and equation~(8) from \citealt{2003A&A...407..957M}).
	Such radio continuum is well above the sensitivity of the 1.3\,cm map from CHWC98,
	the non-detection at the location of source II demands a suppressing mechanism.
	In a stage prior to the \uchii{}, accretion could quench the radio free-free emission 
	\citep{1986ARA&A..24...49Y} and constrain the ionized hydrogen into a small 
	dust-evacuated cavity of radius $R_c \, \sim \, (L_{*}/4 \pi \sigma \, T_{\rm d}^4)^{1/2}$ 
	(see e.g. \citealt{2002hsw..work....3C}).
	Using the same approach as for source III, we determine a lower-limit 
	spectral type between B0 and O9.5 (see Table~\ref{tab:best}).
	For these spectral types, the extent of the ionized region is well traced by the [NeII] line emission.	
	For a luminosity of 10$^5$\,L$_{\sun}$ \citep{1996ApJ...460..914V} and dust sublimation
	temperature $T_{\rm d}\sim1000$\,K, we estimate an $R_c$ of about 50\,AU, which is 0\farcs008 at 
	5.8\,kpc distance. 
	Thus, the typical size of an accretion-quenched \hii{} region would be
	more than two orders of magnitudes smaller than the observed size in the [NeII] line. 
	Therefore, we conclude that no massive star prior to the \uchii{} phase can be responsible 
	for the observed [NeII] emission. 

	In the  third scenario, we assume that the warm dust around one  or more 
	of the \uchii{} regions
	A, B1 or B2 is the source of the mid-infrared flux.
	We will have then to demonstrate that: 
	1) The young massive star ionizing the \uchii{} region
	can maintain an ionized halo detectable in the 
	mid-infrared and showing [NeII] emission at the location of
	source II; 2) The shift between radio and mid-infrared emission can be
	explained by extinction, geometry and/or clumping.
	
	The projected distance between the peak of the mid-infrared source II 
	and the most distant \uchii{} region B1 is 0\farcs8, 0.02\,pc at the distance 
	of G10.47. The spectral types of the  stars ionizing the
	\uchii{} regions A, B1 and B2 are estimated to be B0--O9 or earlier
	from radio measurements (CHWC98).	
	These spectral types agree with our O7 upper limit obtained from
	the ratio of the non-detected and detected forbidden emission lines
	\footnote{
	Estimating the spectral type of the ionizing star following the
	method described in the case of source III would be incorrect. Since we only see
  	part of the mid-infrared emission, the spatial ionization structure of the \uchii{} 
	region should be considered}. 
	The CoStar ionization models predict that the 
	[NeII] emission can extend up to 0.055\,pc from  B0--O9 stars and 
	up to 0.083\,pc from  O9--O8 stars embedded
	in spherically symmetric ionized regions with uniform electron density
	(see Fig.~13 of \citealt{2003ApJ...584..368O}, and text therein).	 
	Thus, the stars powering the \uchii{} regions A, B1 and B2 can indeed provide
	sufficient ionization to explain the observed [NeII] emission.
	
	To explain the apparent shift of the mid-infrared peak from the radio peaks, 
	we consider the following situation:  
	The \uchii{} regions lie in the direction of high line-of-sight extinction and are
	surrounded by warm dust emitting at mid-infrared wavelengths. The peak of the 
	MIR emission would coincide with the \uchii{} regions, but due to the extinction 
	gradient the peak 
	is apparently shifted towards the direction of lower extinction.
	Thus, the MIR-radio continuum offset is the result of extinction and geometry
	of the sources. 		
	In the following more detailed discussion 
	we shall further distinguish between the \uchii{} region A, lying closer to the surface of
	the ammonia molecular core probably in the halo, and the \uchii{} regions B1 and B2 closer to
	the center of the HMC (CHWC98, \citealt{1996A&A...315..565O}). 
	
	First, we consider the case of the two more embedded \uchii{} regions B1 and B2
	and investigate if such embedded sources could be detected in our MIR images.
	CHWC98 estimated B0 spectral types for the stars ionizing the \uchii{} regions B1 and B2
	similarly to the \uchii{} region C.
	We detected MIR emission from dust around the \uchii{} region C (source~III) and we can now
	calculate up to which extinction/column density such emission would be detectable in our
	most sensitive 12.8\,\micron{} broad-band images (see Table~\ref{tab:sens}). 
	By fitting the MIR spectrum of source~III
	we found an extinction of 4\,mag at 9.8\,\micron{} (Table~\ref{tab:best}) 
	that translates into 1.3\,mag at 12.8\,\micron{} assuming the extinction laws from
	\citet{2003ARA&A..41..241D}.  The measured flux of source~III 
	(see Table~\ref{tab:fluxes}) corresponds to
	a de-extincted flux of 1\,Jy in the 12.8\,\micron{} filter.
	Already an extinction/column density 5 times larger than that measured towards source~III 
	reduces the flux density to only 1.5\,mJy, that is below the 1\,$\sigma$ sensitivity of our 
	12.8\,\micron{} images.
	For comparison, the core averaged column density from \citet{1996A&A...315..565O} is more than 70 
	times the column density towards source~III (see Table~\ref{tab:best}). 
	Thus, we conclude that no mid-infrared emission from dust surrounding the \uchii{} regions B1 and B2,
	located close to the core center, could have been detected in our MIR exposures. 	
	We note, however, that the simple approach applied above neglects the clumpiness of the core.
	Should the geometry strongly deviate from the homogenous distribution, 
        radiation from the \uchii{} regions B1 and B2 might also contribute to the MIR flux.

	Proving that the MIR emission comes from dust around the \uchii{} region A is beyond the 
	possibility of this dataset. However, three facts support this scenario.
	First, the column density we measure towards source~II is similar to the averaged column density
	measured by \citet{1996A&A...315..565O} in the halo around the ammonia HMC, where the \uchii{}
	A is probably located.	
	Second, the radio emission estimated from the [NeII] line strength (see above) 
	closely resembles the observed 1.3\,cm flux from the \uchii{} region A (within a factor of two), 
	while showing a larger difference from the 1.3\,cm flux of B1 and B2.	
	Third,  we find a marginal shift in the peak location of source II between the broad-band
	11.9\micron{} filter and the [NeII] narrow-band filter: The [NeII] peak of source II
	is 0\farcs1 N-E of the 11.9\micron{} peak emission, i.e. the peak is closer to the \uchii{} region A.

\subsection{The other mid-infrared sources}\label{others}
	In the large field of view of the images from the second TIMMI2 run we detected seven MIR objects.
	The sources II and III belong to the massive star forming region G10.47
	and have been described in the previous two Sections.
	Source VII is the MIR source located at the largest distance from the HMC 
	in G10.47 and is found to
	coincide with the \uchii{} region G10.46+0.03A within our astrometric accuracy. 
	To investigate the nature of the other four MIR sources, we retrieve their NIR magnitudes from the
	2MASS PSC and we plot their colours in a NIR colour-colour diagram (for the use
	of the colour-colour diagram we refer to \citealt{1992ApJ...393..278L}).
	The colors of main sequence, giant and supergiant stars are taken from 
	\citet{2000.book.....Allen}. 
	
	Source I lies on the right side of the main-sequence stripe, thus indicating NIR excess.
	This excess is thought to originate from hot circumstellar dust. The presence of
	dust around source I is confirmed by our MIR images: Source I appears slightly extended
	in the broad band filters (deconvolved FWHMs of $\sim$\,1\arcsec{} at 11.7 and 11.9\,\micron{} 
	and 0\farcs5 at 12.9\,\micron ) and has an SED slowly increasing towards longer wavelengths
	(see Table~\ref{tab:fluxes}). The continuum-subtracted [NeII] images show no emission. Although
	the further discussion of this object is out of scope of the current paper, we note that
	the lack of the [NeII] and free-free radio emission proves that this star is not a massive 
	(earlier than B0 spectral type) young stellar object.
			 
	Sources IV and V lie within the region of reddened main-sequence stars.
	To determine their spectral types and luminosity classes we adopt the following method:
	I) By shifting these sources back on the colour-colour diagram -- along the reddening vector -- we find the
	possible spectral type/luminosity class combinations. 
	II) From the difference between the old and new location in the colour-colour diagram we 
	estimate the extinction. The distance is given by the comparison of the apparent 
	and the absolute magnitudes taking the derived extinction into account.
	III) As the last test, we compare the 11.9\,\micron{} flux of
	this hypothetical star (at the given distance and extinction) to our measured MIR flux.	
	With this approach we find that sources IV and V are strongly reddened (A$_{\rm V} >20$\,mag). 
	Should they be dwarf stars, they would necessarily lie within a distance of
	80\,pc to fit the apparent brightness
	we observed. However, comparing the A$_{\rm V} >20$\,mag to the mean extinction value of 
	1.9\,V-mag/kpc of the Milky Way disk \citep{1973asqu.book.....A}, 
	we can exclude that these sources are reddened main sequence stars. 
	For MIR IV the most probable nature is that of a G0-G3 supergiant at a distance of
	about 5~kpc, although the measured MIR flux at 11.9\,\micron{} is slightly lower than that estimated from
	a black body approximation ($\sim$44\,mJy) of the photosphere. Similarly, source V is consistent with
	a supergiant of spectral type B6 at a distance of 2.5~kpc and an extinction of 30\,mag in V.
	
\section{Summary} \label{sect:summary}
	We observed the massive star-forming region G10.47 at infrared
	wavelengths in order to study in detail the relationship between the HMC and the
	\uchii{} regions.
	An astrometric accuracy at the subarcsecond level 
	and additional spectroscopy of the two  most
	interesting MIR sources allow us to draw the following main conclusions:  	
\begin{enumerate}
\item We detect extended MIR emission (source II) towards the HMC and the three \uchii{} regions. 
      However, the MIR emission does not coincide with the HMC position nor with the position of any
      of the \uchii{} regions.	
      The most plausible scenario is the one in which the MIR emission of source II originates from
      dust heated  by the \uchii{} region A, one of the three embedded \uchii{} regions
      in the HMC.
      The shift in the peak emission between the ionizing star and the MIR emission
      is interpreted as due to geometrical effect and changes in the extinction. 
      This scenario is supported by the marginal shift of the peak positions found at 11.9\,\micron{}
      and in the [NeII] filter.
\item The \uchii{} region C, which is located outside the HMC, is found
      to have a mid- and a near-infrared counterpart.        
      From our MIR spectroscopy we derive a B0 spectral type for the star ionizing 
      the \uchii{} region C. This spectral type agrees well with that found
      from radio free-free emission (CHWC98).
      The size of the radio free-free emission together with the relatively low column
      density towards the source suggest that the \uchii{} region C could have evolved rapidly because of
      a relatively low density environment.
\item The mid-infrared source VII is coincident with a \uchii{} region in the star forming region
      G10.46+0.03.  
\item The mid-infrared source I shows infrared excess thus indicating the presence of surrounding warm dust.
      The absence of [NeII] and free-free radio emission proves that the star has a spectral type
      not earlier than B0.
\item Combining the 2MASS fluxes with our MIR observations we find that the mid-infrared sources IV and V
      are possibly supergiant stars in the foreground of the massive star-forming region G10.47.
\end{enumerate}      

\begin{acknowledgements}
We wish to thank R. Cesaroni for making available his radio data of G10.47
and N. L. Mart\'{i}n-Hern\'{a}ndez, C. A. Alvarez, E. Puga, F. Wyrowski
and H. Linz for helpful discussion. 
We are grateful to  N. L. Mart\'{i}n-Hern\'{a}ndez for providing the IDL routine 
to calculate the emissivity at the [NeII] line.
We also thank the anonymous referee for useful comments and constructive criticism.
This publication makes use of data products from the Two Micron All
Sky Survey, which is a joint project of the University of Massachusetts
and the Infrared Processing and Analysis Center/California Institute of
Technology, funded by the National Aeronautics and Space Administration
and the National Science Foundation.
\end{acknowledgements}

\bibliographystyle{aa}
\bibliography{lit}
\end{document}